# HABCS*m*: A Hamming Based *t*-way Strategy Based on Hybrid Artificial Bee Colony for Variable Strength Test Sets Generation

**Ammar k Alazzawi[1], Helmi Md Rais[1], Shuib Basri[1], Yazan A. Alsariera[2], Luiz Fernando Capretz[3], Abdullateef Oluwagbemiga Balogun[1] and Abdullahi Abubakar Imam[1]**

[1] Department of Computer and Information Sciences, Universiti Teknologi PETRONAS, Bandar Seri Iskandar 32610, Perak, Malaysia
[2] Department of Computer Science, Northern Border University, Arar 73222, Saudi Arabia
[3] Department of Electrical & Computer Engineering, Western University, 1151 Richmond Street, London, Ontario N6A 5B9, Canada

Corresponding author: Ammar K Alazzawi (e-mail: ammar_16000020@utp.edu.my).

## Abstract

Search-based software engineering that involves the deployment of meta-heuristics in applicable software processes has been gaining wide attention. Recently, researchers have been advocating the adoption of meta-heuristic algorithms for *t*-way testing strategies (where *t* points the interaction strength among parameters). Although helpful, no single meta-heuristic based *t*-way strategy can claim dominance over its counterparts. For this reason, the hybridization of meta-heuristic algorithms can help to ascertain the search capabilities of each by compensating for the limitations of one algorithm with the strength of others. Consequently, a new meta-heuristic based *t*-way strategy called Hybrid Artificial Bee Colony (HABCS*m*) strategy, based on merging the advantages of the Artificial Bee Colony (ABC) algorithm with the advantages of a Particle Swarm Optimization (PSO) algorithm is proposed in this paper. HABCS*m* is the first *t*-way strategy to adopt Hybrid Artificial Bee Colony (HABC) algorithm with Hamming distance as its core method for generating a final test set and the first to adopt the Hamming distance as the final selection criterion for enhancing the exploration of new solutions. The experimental results demonstrate that HABCS*m* provides superior competitive performance over its counterparts. Therefore, this finding contributes to the field of software testing by minimizing the number of test cases required for test execution.

## 1. Introduction

One of the most important processes in any Software-Development Life Cycle (SDLC) is Software Testing [1]. Testing is important for identifying the parts of software that are not performing as expected (i.e. to prevent the defects or bugs) and for guaranteeing the software comes up with all of its specifications. This process is expensive, due to the amount of time required to implement the test sets. For this reason, many strategies are proposed for generating an efficient test set to test any software system. These include Equivalence Partitioning, Boundary Value Analysis, and Decision Tables etc. While these strategies are useful for specific problems, they have not effectively addressed the problems related to very large parameters that are due to interaction. Accordingly, researchers have made great efforts over the past few years to minimize the size of the test sets in order to lessen the cost of the software testing process and to preserve their ability to detect defects.

There are a plethora of contributions available in the literature, whose main purpose is to minimize the number of test cases by removing repeated test cases [2, 3]. Therefore, the execution of test cases will call for less effort [4]. One of the most effective approaches to lessening the number of test cases is Combinatorial Testing (CT) [5], otherwise known as Combinatorial Interaction Testing (CIT). CT has recently been used to test software systems in multiple areas of interest. For example, the number of configurable parameters of the real software systems is large. As a result, it is impossible to test all configurations because of limited resources. The empirical evidence shows CT is achieving very high fault coverage of effectively detecting faults by implementing a small number of test cases on the system. One of the most efficient and effective kinds of CT is *t*-way testing (where *t* is the degree of interaction strength) for feasible solutions based on the failures caused by the value of *t* interaction parameters in the configuration system (inputs).

Many *t*-way strategies are categorized to algebraic and computational based strategies. Algebraic strategies are based on exploiting the same mathematical functions as orthogonal arrays (OA) [6]. Despite the fact that these strategies are fast and generate optimal solutions, but they do not support the high interaction strength of some configurations systems because of the restrictions that exist. To ease and overcome these restrictions, computational based strategies have been proposed to support the arbitrary configuration. These strategies generate high-quality solutions [7]. Recently, the efforts of researchers have been focused on meta-heuristic algorithms as a foundation for *t*-way strategies. These include Bat Algorithm [8-13], Artificial Bee Colony [14-17] and Kidney Algorithm [18], to name just a few. However, it is impossible to find meta-heuristic algorithms to compensate for other existing algorithms for all optimization problems. Consequently, hybridization is the most effective and efficient method for improving the performance of *t*-way strategies, based on merging and exploiting the strengths of two or more algorithms.

One of the most effective meta-heuristic algorithms is an Artificial Bee Colony (ABC) algorithm [19], which has been applied successfully to various optimization problems. Several studies have documented the performance of the ABC algorithm on optimization problems [20-23]. However, ABC algorithm has suffered from problems with convergence speed, simple operation of solution development and weak information sharing activity. For this reason, many researchers have proposed variants of the ABC algorithm, either by modifying it based on hybridizations or by improving the original ABC, such as Bee Swarm Optimization (BSO) [24], gbest-guided ABC (GABC) [25], hybrid ACO–ABC [26] and others. In this paper, a new meta-heuristic based *t*-way strategy called Hybrid Artificial Bee Colony (HABCS*m*) strategy is being proposed. This strategy is based on merging the advantages of the Artificial Bee Colony (ABC) algorithm with the advantages of a Particle Swarm Optimization (PSO) algorithm to improve the performance of the original ABC algorithm. HABCS*m* is the first *t*-way strategy to adopt Hybrid Artificial Bee Colony (HABC) algorithm as its core application for generating a final test set and the first to adopt the Hamming distance as the final selection criterion for enhancing the exploration of new solutions.

The paper is presented as follows: Section 2 presents the theoretical concept of covering array notation. Followed by related works in section 3. Section 4 provides an overview of the Hybrid Artificial Bee Colony algorithm. Section 5 discusses the phases of our proposed strategy. Section 6 shows the results of tuning experiments. Section 7 shows the results of our benchmarking experiments. Finally, Section 7 provides the overall concluding remarks along with future work.

## 2. Main Concept of Covering Array Notation

Mathematically, Covering Array (CA) is a widely reputable mathematical theme, whereas CT depends on the CA for theoretical test set size generation [27]. CA is available as another practical alternative to the previous (or the oldest) mathematical theme, namely Orthogonal Array (OA), which is used for the purpose of statistical experiments [28]. In general, each of the software being tested (SUT) includes several items such as the parameters ($P$) and value ($V$) of each parameter and the interaction strength degree ($t$).

An orthogonal array utilization is limited in the software testing field because it requires that all parameters have the same value number, and that each pair of values be covered the same number of times [6, 29]. Therefore, OA has difficulty producing the test set, and frequently it is quite large. However, on the positive side, OA is characterized by an ability to easily recognize the combination which is responsible for causing a failure [30]. To ease and overcome the constraints of OA, CA and Mixed CA (MCA) have been proposed.

CA is defined by CA ($N$, $t$, $v^P$) [1], as a two-dimensional array, where $N$ indicates the number of test cases (i.e. rows), $P$ indicates the parameter's number (i.e. columns), $V$ indicates the value of each parameter, and $t$ indicates the degree of interaction strength . The optimal CA includes the minimum number of test cases, called test sets. CA is an array $N$ x $P$, where $N$ indicates the number of test cases (i.e. rows), and $P$ indicates the parameter's (i.e. columns). For example, this CA (16, 2, $4^6$) has an interaction strength degree of 2 and the parameter's number is 6, and these are linked with 4 values each. In cases where each parameter has a different value, this is called Mixed Covering Array (MCA) [28], which is defined by MCA ($N$, $t$, $v_1^{P1}$, $v_2^{P2}$, ..., $v_i^{Pi}$). Consider MCA (9, 2, $3^6\ 2^4$) as an example. The MCA interaction strength degree is 2 and the parameter's number is 6, and these are linked with 3 values each and 4 are linked with 2 values each.

With the rapid improvement in information technology (IT) and complexity in modern software structure, a Variable Strength Covering Array (VSCA) alongside CA and MCA is been implemented. A compelling difference was observed in the interaction strength amongst such software structure parameters [31].VSCA is defined by VSCA ($N$; $t$, $v^P$, ($CA_i$)), where $CA_i$ is the subdivision of the main space set with a distinct interaction degree.

## 3. Related Work

In general, existing state-of-the-art strategies for generating the optimal or best test set sizes are divided into two approaches: Algebraic and Computational [17, 32]. Mathematicians have typically utilized the Algebraic approach, where a test set is constructed based on the construction of OA. The OA is derived from the extension of mathematical functions [33]. Although Algebraic based strategies are fast, their utilization in the CT field is restricted to small configurations (e.g. all parameters must have the same exact value number) [2, 34]. Strategies that have utilized the Algebraic approach are Test Configuration (TConfig), Orthogonal Latin Squares (OLS) [6], CA [1] and MCA [28]. The second strategy is the Computational approach, which is based on greedy algorithms for generating test set size to cover the maximum number of combinations. This approach is classified as one-parameter-at-a-time (OPAT) or one-test-at-a-time (OTAT) approach to generate the final test set size.

The one-parameter-at-a-time approach begins by generating a complete test set for the first two parameters and expands this process horizontally, and occasionally vertically, by adding one parameter at a time per iteration to ensure coverage of all parameters [35]. The known strategies that adopted OPAT approach are the in-parameter-order (IPO) strategy [35] and its improvements such as IPAD2 [36], IPOG-D [37] and IPOG [38], to name just a few. One-test-at-a-time strategies begin by generating a complete test case at a time per iteration to ensure that all interaction components are covered. The first strategy to adopt this approach is Automatic Efficient Test Generator (AETG) [39], and this was followed by many other strategies such as Jenny [40], WHITCH [41], TConfig [27], Test Case Generation (TCG) [42], Deterministic Density Algorithm (DDA) [43], Classification-Tree Editor eXtended

Logics (CTE-XL) [44], Pairwise Independent Combinatorial Testing (PICT) [45] and Intersection Residual Pair Set Strategy (IRPS) [46].

Recently, researchers began focusing on meta-heuristic algorithms as a main algorithm for t-way test set generation strategies such as Bat Algorithm (BA) [8-13], Hill Climbing (HC) [28], Simulated Annealing (SA) [27, 28, 31, 41], Tabu Search (TS) [47, 48], Artificial Bee Colony (ABC) [14-16], Particle Swarm Optimization (PSO) [49], Cuckoo Search (CS) [50], Ant Colony Algorithm (ACA) [51], Genetic Algorithm (GA) [52], Kidney Algorithm (KA) [18] and others. The most essential algorithm used for a 2-way test set generation ( also termed pairwise test set generation) is HC [28]. However, the initial search of HC is sensitive. For this reason, it is apt to be restricted in the local optima. Another algorithm used successfully for pairwise test set generation is TS [47, 48]. Due to constraints of HC and TC, and lack of support for a large the interaction strength more than t=2. Stardom proposed SA as an improvement of HC [27, 28, 31, 41]. SA is used for t-way test set generation and to support higher interaction strength $t \leq 3$, which is dissimilar to HC and TS. The experimental results have shown that SA outperformed both HC and TC, especially for higher strengths $t \leq 3$. In early studies, the Meta-heuristic algorithms are adopted to generate uniform CA and VSCA, such as GA [52] and ACA [51], while GA begins searching in all available positions in order to find the solutions randomly. Therefore, all solutions subjected to repeat cycles of processes such as mutation and selection, crossover in order to mimic natural selection of biological evolution, unlike SA, HS and TS. In contrast, ACA was inspired by Ant Colony behaviour in order to find the best food paths. GA and ACA often are not stuck in local optimum compared to HC, TS and SA. Moreover, ACA and SA were improved to support a VSCA up to $t \leq 3$.

Bestoun et al. [53] proposed the Particle Swarm Test Generator (PSTG) strategy for t-way test set generation and variable strength PSTG (VS-PSTG) strategy [54] utilizing PSO algorithm. Both the PSTG and VS-PSTG strategies are based on and inspired by the behaviour of birds in a swarm when searching for food. Several parameters of PSO have been tuned perfectly to find the best particles or solutions. PSO was then extended for a better solution called discrete PSO (DPSO) [55], whereby the candidate solution is represented by a particle's position. The results of these experiments show that DPSO outperformed the original PSO by producing optimal or near-optimal solutions. PSO outperformed GA and ACA in terms of support for high-interaction strengths *t*=6. However, GA and ACA had relatively better computation times compared to PSO [56]. HS strategy as proposed by Alsewari et al. [57] is similar to PSTG and VS-PSTG for t-way test set generation. HS strategy is inspired by the behaviour of a skilled musician who is creating good music. To explore the search spaces efficiently, HS strategy used elitism selection that was used in GA, for intensification and diversification. Furthermore, mathematical equations are used to find relatively better solutions and a probabilistic gradient to select the current solution neighbour.

## 4. Hybrid Artificial Bee Colony Algorithm

In 2005, Karaboge proposed the Artificial Bee Colony (ABC) algorithm to solve an optimization problem [19]. ABC algorithm is considered one of the most efficient meta-heuristics algorithms and can be applied in many different areas of optimization problems, such as the combinatorial optimization problem. ABC algorithm mimics the foraging behaviour of honeybees inside the hive. This algorithm consists of three kinds of bee, where each bee executes a different task in order to find the best food source (i.e. solution). These three kinds of bees are employed, onlooker and scout bees. Employed bees begin to explore the entire environment in order to find food sources with the maximum amount of nectar. Then, they communicate the information about food sources (i.e. profitability, distance, direction) with other kinds of bees that are waiting inside the hive (i.e. Onlooker bee). Onlooker bees take the information that is given by the employed bee in order to make the decision to choose the food with the maximum amount of nectar. The Scout bee works to detect a new or better food source than the existing food source by randomly searching the environment again.

In the past several years, many meta-heuristic algorithms such as Artificial Bee Colony (ABC) [19] have achieved significant success in solving optimization problems. ABC algorithm works well and is appropriate for several optimization problems. However, ABC algorithm has several weaknesses as well. These include easily becoming stuck in a local optimum when handling some complex problems (i.e. exploitation process), simple operation of solution development and weak performance of information sharing activity. For this reason, many researchers have proposed using variants of ABC algorithm, either by improving the original ABC program or by modifying it based on hybridization. This would help to even out the exploration and exploitation processes, which would then help to avoid becoming stuck in the local optimum and create better convergence, as can be seen in Bee Swarm Optimization (BSO) [24], gbest-guided ABC (GABC) [25], hybrid ACO–ABC [26] and others.

Every meta-heuristic algorithm utilizes a different strategy of exploration and exploitation for optimization problems. Recently, new research on optimization problems has been attracting attention. It uses a hybrid algorithm to overcome the problem of the poor exploitation mechanism of ABC. In this paper, a new hybrid metaheuristic algorithm is proposed based on merging the advantages of ABC and PSO. Thus, this paper proposes a new algorithm called Hybrid Artificial Bee Colony (HABC) algorithm, which utilizes the exploitation ability of PSO (i.e. local search process) and the exploration ability of ABC (i.e. global search process). While the original PSO has considerable exploitation ability and a fast convergence speed [58], it has poor exploration ability. By contrast, the original ABC, has great exploration ability but poor exploitation mechanism abilities [25].

Therefore, this paper proposes a Hybrid Artificial Bee Colony (HABC) algorithm that adopts particle characteristics of the Particle Swarm Optimization (PSO), such as solution development mechanism, which represents the main operator for exploitation and exploration. The solution development mechanism of Particle Swarm Optimization (PSO) algorithm (i.e. the exploitation ability process) is unique and different from an exploitation mechanism of the original Artificial Bee Colony (ABC) algorithm. The Particle Swarm Optimization (PSO) algorithm solution development mechanism that formulated in Equation (2). In each iteration, the velocities and positions of the particles are updated by formulated Equation (2) and Equation (3) respectively.

Based on the assertions stated above, the Hybrid Artificial Bee Colony (HABC) algorithm consists of five steps: initialization phase, employed bees' phase Based on the PSO development mechanism, calculation of the probability of selecting a food source, onlooker bees' phase and scout bees' phase. These phases are explained as follows:

At the beginning of HABC algorithm, the initialization phase begins randomly generate a food source (i.e., a potential solution for a certain optimization problem) by formulating an equation (1). Each food source (i.e. solution) is indicated as $X_i = (X_{i1}, X_{i2}, X_{i3},..., X_{iD})$, $i \in \{1, 2,..., SN\}$, where $SN$ indicates the swarm size. Where D random number of dimension index. The swarm size is equal to the number of food sources and $Xmin$ and $Xmax$ are the lower and upper bounds of the food source position respectively.

$$x_{ij} = x_{min,j} + rand\,(0,\,1)\,(x_{max,j} - x_{min,j}) \qquad (1)$$

In the employed bee phase, the solution development mechanisms of PSO (i.e. exploitation ability process) are performed in order to generate new food sources in the search space using equation (2) and (3) instead of the original equation of employed bee in ABC algorithm. First, each food source needs to be checked in each iteration, and **pbest** records the updated status of each one.

$$V_{i,d}^{t+1} = W^t * V_{i,d}^t + C_1^t * r_1 * (pbest_{i,d}^t - X_{i,d}^t) + C_2^t * r_2 * (gbest_{i,d}^t - X_{i,d}^t) \qquad (2)$$

$$y_i^{t+1} = y_i^t + V_i^{t+1} \qquad (3)$$

Where consists of several important parameters such as Velocity parameter, indicated as $V_i = (V_{i1}, V_{i2}, V_{i3},...,V_{iD})$, where velocity is used to control the improvement solutions, the current position of $i$ particle indicated as $y_i = (y_{i1}, y_{i2}, y_{i3},...,y_{iD})$, and the previous best position of particle, indicated as $pbest_i = (pbest_{i1}, pbest_{i2}, pbest_{i3},...,pbest_{iD})$. The best position determined among all the previous best positions is indicated as $gbest = (gbest_{i1}, gbest_{i2}, gbest_{i3},..., gbest_{iD})$. Where both **C1** and **C2** are two positive constants, they indicate the relative influence of the cognition and social components. **W** indicates the inertia weight that tools up balance between local exploitation and global exploration mechanism. Both $r_1$ and $r_2$ indicate the real random value from 0 to 1. The particle's velocity on each dimension is clamped to the range $[-Vmax, Vmax]$.

Then, a calculation is done of the probability of selecting a food source by adopting a greedy selection mechanism to check the better fitness value of the old and new food source using equation (3). When all criteria have been exhausted, the **pbest** value will stop the updating process (i.e. solution development mechanism of PSO in the employed bee phase will be terminated) and the HABC algorithm performs the onlooker bee phase.

$$fit_i = \begin{cases} \frac{1}{1+f_i}, & if\ f_i \geq 0 \\ 1 + |f_i|, & if\ f_i < 0 \end{cases} \qquad (4)$$

In the onlooker bee phase, the probability ($Pi$) of food source selection is dependent on the amount of nectar ($X_i$), where the employed bee propagates the information about the food sources, to the onlooker bee on the dancing area inside the hive. Then the onlooker bee selects the best food source with a maximum amount of nectar by a formulated equation (5). After the onlooker bee probability selection process is finished, the onlooker bee starts a new search for a new food source, while trying to improve the food source of the employed bee by using a formulated equation (2).

$$Pi = \frac{fit_i}{\sum_{n=1}^{SN} fit_n} \qquad (5)$$

Upon completion of both the employed and onlooker bee phases, the search process continues for further improving the food sources. In the scout bee phase, the algorithm will check the search space for further improvements using a number of trial cycles called "*limit*" that are formulated by equation (6). The scout bee selects randomly from best previous solution **pbest** and generates a new food source using equation (2). Once the fitness value of a new food source better or equal to the previous best solution has been identified, the new one will be selected. Otherwise, the previous best solution is retained.

$$limit = c.ne\,.D \qquad (6)$$

Where **C** is a constant coefficient with a recommended value of 0.5 or 1, **Ne** indicates the number of unemployed bees (i.e. onlooker and scout bee), and **D** indicates the food source positions. The main steps of HABC algorithm are given below in Figure 1.

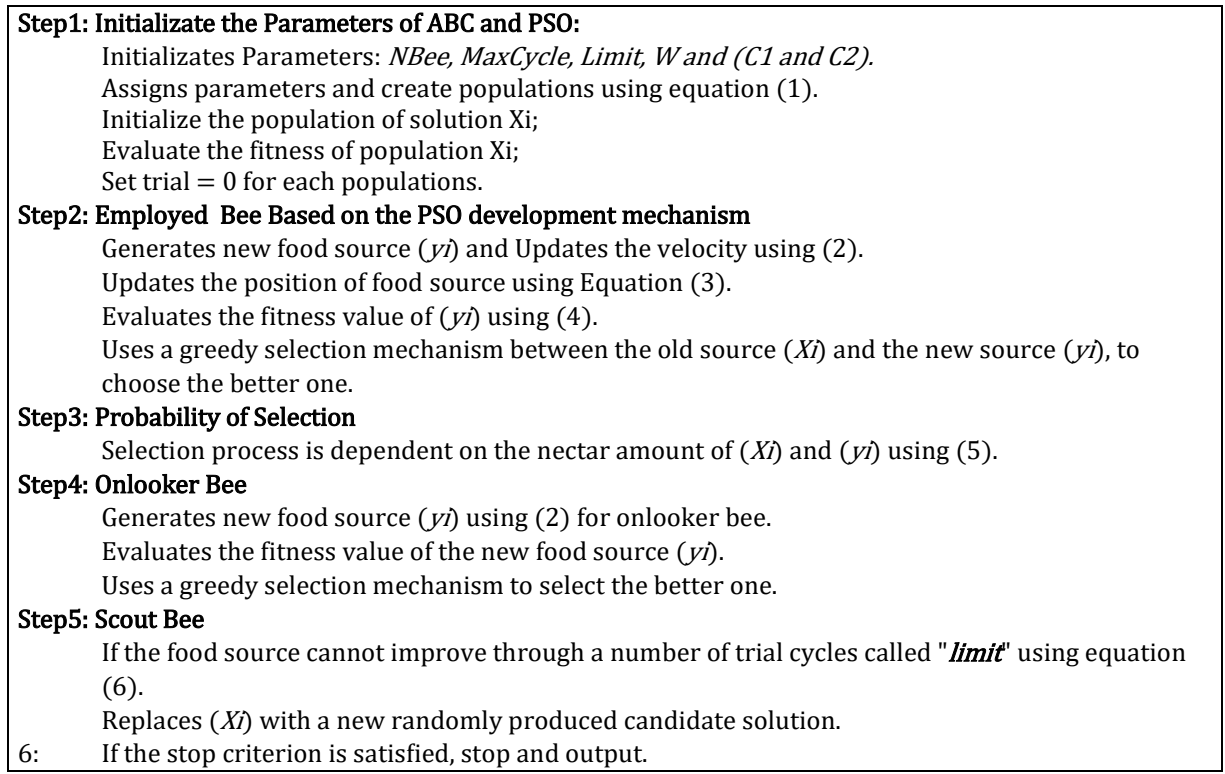

**Figure 1:** The main steps of Hybrid Artificial Bee Colony algorithm (HABC).

## 5. Hybrid Artificial Bee Colony (HABCSm) Strategy

In this research, a new strategy is proposed for generating the optimal test set. The HABCS*m* strategy undergoes three phases during the construction of the mixed and variable strength *t*-way test set generation. The phases illustrated in Figure 2 processed as follows:

**Phase 1.** The HABCS*m* input analysis algorithm exploits the input analyser.

**Phase 2.** The HABCS*m* interaction generation algorithm, which is based on the Combination *t*-tuple Sets (CTS) generator and Interaction Elements Tuple (IET) generator, generates the required *t*-tuples.

**Phase 3.** The HABCS*m* test set generation, which is based on exploiting the HABC algorithm as the core algorithm and the Hamming distance selection criterion for mixed and variable strength *t*-way test set generation.

These phases will be discussed further and elaborated on in the next three sub-sections.

## A. Input Analysis

In this section, the input (CA, MCA and VSCA) processing is divided into two main processes: processing the input components (i.e. parameters and their values), and representing these components using numerical values. The first process starts by receiving the input and processing the input components to a set of pre-defined variables in the memory (i.e. interaction strength (*t*), parameters (*P*) and their values (*v*)).

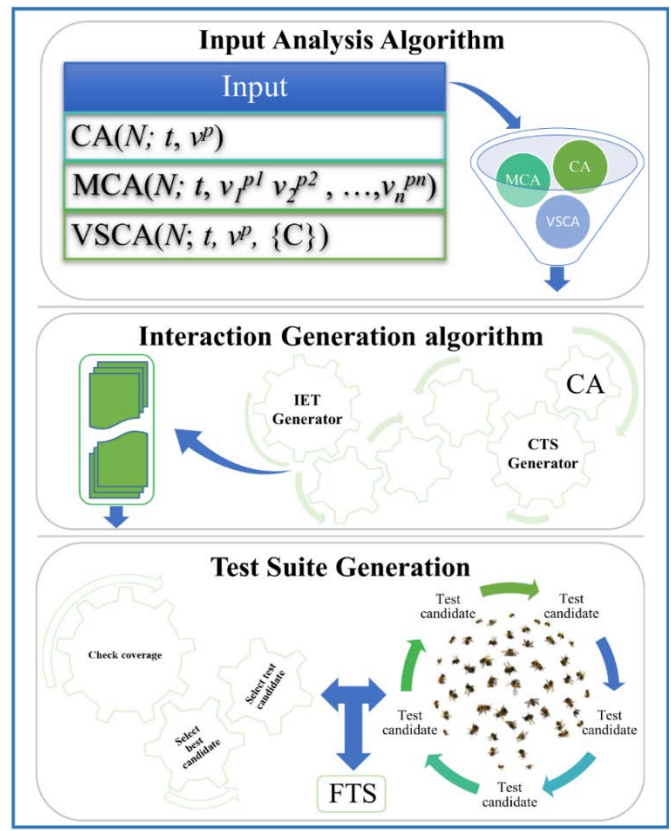

**Figure 2.** An Overview of HABCS*m* strategy.

**B. Interaction Generation Algorithm**

This section includes the generation of the Combination *t*-tuple Sets (CTS), and the Interaction Element Tuples (IET), based on the parameters and their value for the configuration system. HABCS*m* employs a new-generation approach that reversely generates the Binary Element Set (BES). BES represents the evaluation of each test case (i.e. the fitness value of each test case provided by HABC algorithm). This phase includes two algorithms; the CTS and IET generators.

The first algorithm (CTS generator) generates the Combination *t*-tuple sets (CTS), and after completing the input analysis, the numerical value will be generated. This numerical value is for the generation of the element combination, which is the basis of CTS. The generation starts as soon as it receives the numerical value and the interaction strength (*t*) value from the input analysis phase.

The second algorithm in this phase is the Interaction Elements Tuple (IET), which is used to construct all the interaction elements, based on the represented binary element for all the combinations in CTS. The two processes of generating IET and BES are combined into one algorithm to reduce the complexity of the generation process (i.e. the generation of all the possible binary elements and the selection of the binary elements that seems to be inefficient when done separately). IET generation starts with the selection of all the combinations in the CTS and explores them until all the interaction elements (IE) are generated for each combination.

**C. Test Set Generation Algorithm**

The HABC algorithm can improve the solution quality because of the global and local search behaviour it implements. Therefore, the HABC algorithm is employed as a search engine to calculate the fitness (coverage or weight) of the randomly generated test candidates for the proposed strategy. To achieve a minimum test set optimization process, the test cases need to effectively and greedily cover all the *t*-way tuples, at one time, if possible. The HABC algorithm has conventionally developed on the assumption of food source numbers within populations. In applying this algorithm for interaction testing, we assume that the test candidates are food sources and that each source has its own possible solution (fitness) for the targeted problem. The HABC algorithm search process in the proposed strategy provides the best global optimum (or optimum test candidate that has the highest coverage of the *t*-tuples element values) based on the number of BEs involved. This optimum test candidate indicates the solution quality in terms of the best food source.

Unlike the standard HABC algorithm, the HABCS*m* strategy introduces the Hamming distance classifier (See equation 7) to decide the final set. Specifically, the Hamming distance classifier measures two rows of (best) test cases (as string) based on the number of values in which they differ when there is a tie situation as far as the quality of the test cases are concerned. It is the test case that is at the greatest distance that will be finally selected by the Hamming distance classifier to ensure sufficient exploration and exploitation of the search space.

$$Hamming\ distance\ (d(v)) = \sum |v_i - TC_n| \qquad (7)$$

### 5.1 Parameter Tuning of HABCS*m* Strategy

In order to ensure the most optimal results for the HABCS*m* strategy with regards the test set size, this section elaborates on the process of the tuning variables for the HABCS*m* strategy. The control variables for the HABC algorithm have to be tuned based on the test set generation problem. For tuning purposes, a well-known test system (covering array) that involves a CA (N; 2, $5^7$) is employed. The justification for adopting this configuration for the tuning process originates from the use of the same CA to tune many of the existing *t*-way strategies [54, 57, 59]. The process of tuning HABCS*m* is based on 20 runs for the specified CA with different variables settings. The HABCS*m* has six main control variables: {Bee population size (*NBee*), Food source= *NBee*/2, *Limit*, Maximum cycle number (*MCN*), Self-confidence factor and Swarm confidence factor (*C1* & *C2*), and Inertia weight factor (*W*)} that control the sizes of the test sets obtained. The size and average of the final test sets for the 20 runs were recorded. Then, the results of the tuned variables are analysed to find the settings that fit the minimum size and average of the final test sets as shown in Figure 3 and Figure 4. The six variables are executed for all the possible selected settings.

The first variable is bee population size (*NBee*), which indicates the number of bees involved in the test set generation. This variable controls the randomly initialized test candidates in the memory. When the number of test candidates increases, the possibility of finding a better solution (i.e. a test candidate with maximum interaction elements coverage) improves. The initial values of all control parameters need to be tuned: The Bee population sizes (*NBee*) are in the range of [4-9 & 10-100], Food source= *NBee*/2, *Limit*=100, Maximum cycle number (*MCN*) = [10-100 & 200-2000], Self-confidence factor and Swarm confidence factor (*C1* & *C2*) = are set between 0.1 and 2.1, and Inertia weight factor (*W*) = are set between 0.1 and 1.0. Thus, a set of values is selected in this range and all variables are selected to fit the test set generation.

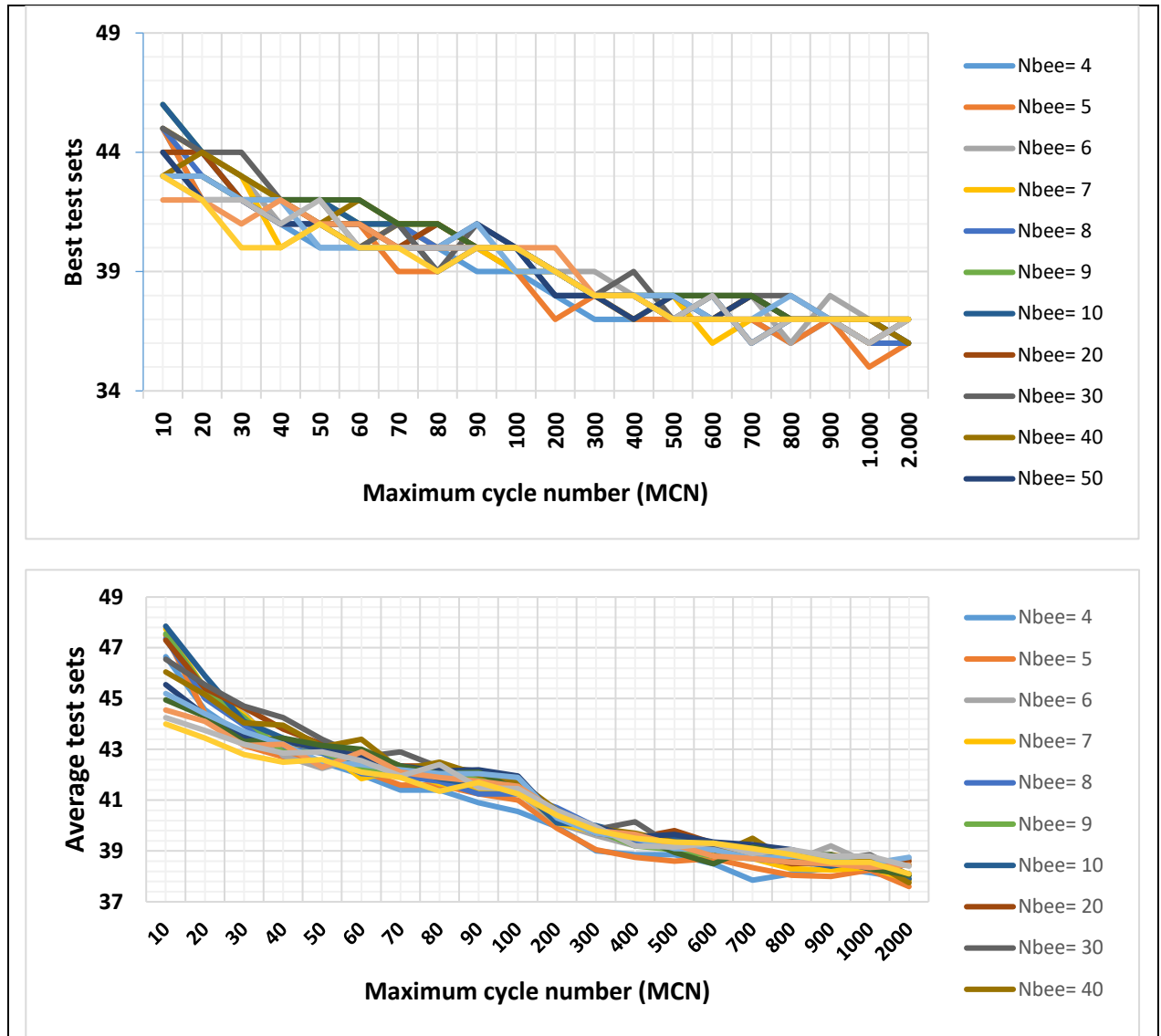

**Figure 3:** Best and Average Test Suite for CA (N; 2, $5^7$).

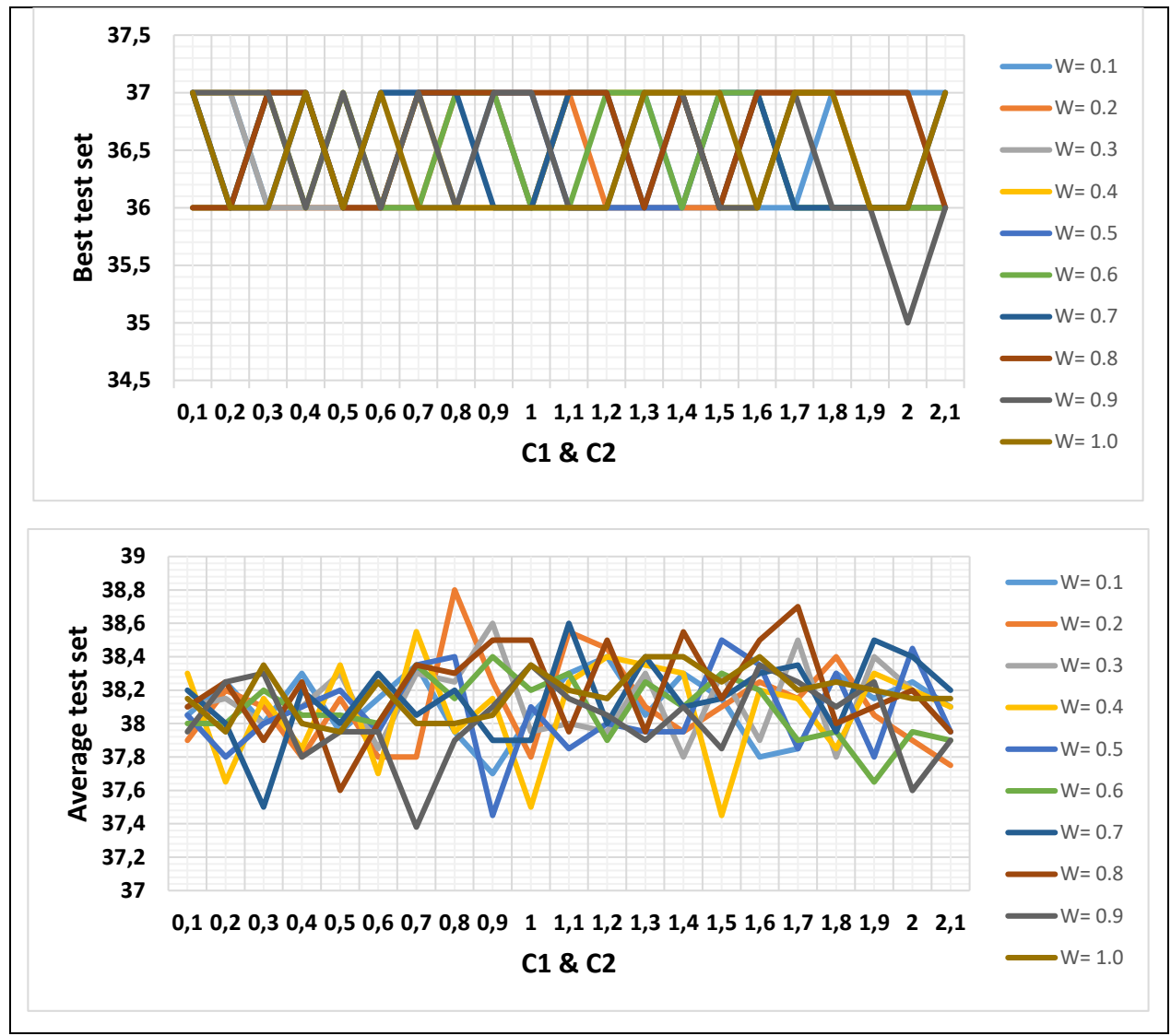

**Figure 4:** Best and Average Test Suite for CA (N; 2, $5^7$).

Based on the empirical experiments and the results of this study as shown in Figure 3, the differences in the test set slightly decreased when the bee population decreased in size. The same observation applies for Maximum cycle number (MCN), where clearly increasing the MCN can produce a good result. The HABCS*m* strategy produces poor outcomes when the number of MCN < 100. On the other hand, the best outcomes are obtained when the MCN is more than 100 in simulations. Finally, the best test sets have been recorded when (*NBees*) = 5 and the (*MCN*) = 1000. In sum, the HABCS*m* strategy generates the best test set size when the MCN is greater than 700.

After determining the best value for (*NBees*) and the (*MCN*), it's important to determine the Self-confidence factor and Swarm confidence factor (*C1 & C2*) and Inertia weight factor (*W*). During the tuning, the value of (*NBees*) and the (*MCN*) are fixed to the values deduced earlier, (*NBees*) = 5 and the (*MCN*) = 1000. Different values of (*C1 & C2*) are tested by fixing the (*C1 & C2*), and change the (*W*) value, and then same process will perform for W (i.e. performing the reverse experiment). As shown in Figure 4 (C1 & C2) and (W) have major impact on the test size generation. The HABCSm strategy has poor results with range of (C1 & C2) greater than 0.1 and smaller than 1.2. Whereas HABCSm strategy performs well in case of (C1, C2) greater than 1.3 when W increases more than 0.5. Based on the aforementioned discussion, the best test sets have been recorded when (C1 & C2) = 2.0 and (W) = 0.9.

## 6. Result and Discussion

This section shows the experiments of HABCS*m* strategy. The experiments were conducted to evaluate and compare the efficiency of the proposed HABCS*m* with our previous work [14-16, 60-65], based on the original Artificial Bee Colony strategy and Hybrid Artificial Bee Colony strategy, as well as with existing published work as adopted from [2, 4, 17, 50, 57, 66]. Whereas implementation times were neglected due to variances in parameter settings (e.g., SA relies on the Iteration, Cooling schedule, and Starting temperature, while the ABCS relies on the Bee population size, Food source number, Limit and Maximum cycle number) and running platform environment (e.g., the implementation language and data structure).

The results of the experiments were performed on the Windows 7 (OS) desktop computer with 3.40 GHz Xeon (R) CPU E3 and 8GB RAM. The Java language JDK 1.8 was used to code and implement the HABCS*m*. Due to the randomization characteristics of the proposed HABCS*m* strategy and the evidence in the literature, the experiments were run 20 independent times for each configuration system to get the best results. The best result was reported as bold and shaded cells to give a better indication. Tables 1 to 7 highlight our experimental results. The selected sets of benchmarked experiments are as follows:

Referring to experimental sets 1, 2 and 3, the HABCS*m* strategy displays better performance and efficiency than the ABC variants (e.g., ABCS and HABC). When the HABCS*m* strategy is compared to Jenny, TConfig, PICT, IPOG-D, IPOG, DPSO, PSTG, CS, GS, HSS, ABCS and HABC strategies. For experimental set 1 in Table 1, the HABCS*m* generates the best test set sizes (or most minimum 17 out of 40 entries). Clearly, the HABCS*m* contributes by providing 6 out of the 17 entries with new minimal test set sizes. Whereas DPSO outperformed all the existing strategies to produce the best test set sizes (or most minimum of 23 out of 40 entries). The other strategies perform well at low interaction strengths ($t \leq 3$), while TConfig, and PICT have not reported significant or minimum test set size for any entries.

Table 1. Test set size for CA (N; $t$, $3^P$) with $2 \leq t \leq 11$ and $3 \leq P \leq 12$.

| | | Pure computation strategies | | | | | AI-based strategies | | | | | | | | |
|---|---|---|---|---|---|---|---|---|---|---|---|---|---|---|---|
| $t$ | P | Jenny Best | TConfig Best | PICT Best | IPOG-D Best | IPOG Best | DPSO Best | PSTG Best | CS Best | GS Best | HSS Best | ABCS Best | HABC Best | HABCS*m* Best | HABCS*m* Avg.Size |
| 2 | 3 | **9** | 10 | 10 | 15 | **9** | **9** | **9** | **9** | **9** | **9** | **9** | **9** | **9** | 9.80000 |
| | 4 | 13 | 10 | 13 | 15 | **9** | **9** | **9** | **9** | **9** | **9** | **9** | **9** | **9** | 10.5500 |
| | 5 | 14 | 14 | 13 | 15 | 15 | **11** | 12 | **11** | **11** | 12 | **11** | 12 | **11** | 12.6500 |
| | 6 | 15 | 15 | 14 | 15 | 15 | 14 | **13** | **13** | **13** | **13** | **13** | **13** | **13** | 14.2000 |
| | 7 | 16 | 15 | 16 | 15 | 15 | **14** | 15 | **14** | **14** | 15 | 15 | 15 | **14** | 15.3000 |
| | 8 | 17 | 17 | 16 | **15** | **15** | **15** | **15** | **15** | **15** | **15** | **15** | **15** | **15** | 16.2000 |
| | 9 | 18 | 17 | 17 | **15** | **15** | **15** | 17 | 16 | **15** | 17 | 16 | 16 | **15** | 16.8000 |
| | 10 | 19 | 17 | 18 | 21 | **15*** | 16 | 17 | 17 | 16 | 17 | 17 | 17 | 16 | 17.2000 |
| | 11 | 17 | 20 | 18 | 21 | 17 | 17 | 17 | 18 | **16** | 17 | 17 | 18 | 17 | 18.1500 |
| | 12 | 19 | 20 | 19 | 21 | 21 | **16** | 18 | 18 | **16** | 18 | 18 | 18 | 18 | 19.4500 |
| 3 | 4 | 34 | 32 | 34 | **27** | 32 | **27** | **27** | 28 | **27** | 30 | **27** | **27** | **27** | 29.7500 |
| | 5 | 40 | 40 | 43 | 45 | 41 | 41 | 39 | **38** | **38** | 39 | **38** | 39 | 39 | 43.0000 |
| | 6 | 51 | 48 | 48 | 45 | 46 | **33** | 45 | 43 | 43 | 45 | 44 | 43 | 43 | 45.5500 |
| | 7 | 51 | 55 | 51 | 50 | 55 | 48 | 50 | 48 | 49 | 50 | 49 | 47 | **46** | 51.2000 |
| | 8 | 58 | 58 | 59 | 50 | 56 | 52 | 54 | 53 | 54 | 54 | 54 | 53 | **45** | 54.3500 |
| | 9 | 62 | 64 | 63 | 71 | 63 | **56** | 58 | 58 | 58 | 59 | 58 | **56** | **56** | 58.8500 |
| | 10 | 65 | 68 | 65 | 71 | 66 | **59** | 62 | 62 | 61 | 62 | 62 | 61 | 61 | 63.5000 |
| | 11 | 65 | 72 | 70 | 76 | 70 | **63** | 64 | 66 | **63** | 66 | 66 | 68 | 65 | 67.0000 |
| | 12 | 68 | 77 | 72 | 76 | 73 | **65** | 67 | 70 | 67 | 67 | 70 | 72 | 68 | 70.2000 |
| 4 | 5 | 109 | 97 | 100 | 162 | 97 | **81** | 96 | 94 | 90 | 94 | 98 | **81** | **81** | 91.9500 |
| | 6 | 140 | 141 | 142 | 162 | 141 | 131 | 133 | 132 | **129** | 132 | 135 | 134 | 132 | 137.100 |
| | 7 | 169 | 166 | 168 | 226 | 167 | 150 | 155 | 154 | 153 | 154 | 157 | 155 | **149** | 158.250 |
| | 8 | 187 | 190 | 189 | 226 | 192 | 171 | 175 | 173 | 173 | 174 | 179 | 177 | **159** | 179.250 |
| | 9 | 206 | 213 | 211 | 260 | 210 | 187 | 195 | 195 | 194 | 195 | 197 | 196 | **185** | 197.600 |
| | 10 | 221 | 235 | 231 | 278 | 233 | **206** | 210 | 211 | 209 | 212 | 215 | 217 | 212 | 215.150 |
| | 11 | 236 | 258 | 249 | 332 | 251 | **221** | 222 | 229 | 223 | 223 | 234 | 237 | 229 | 232.200 |
| | 12 | 252 | 272 | 269 | 332 | 272 | 237 | 244 | 253 | **236** | 244 | 251 | 257 | 246 | 244.550 |
| 5 | 6 | 348 | 305 | 310 | 386 | 305 | 244 | 312 | 304 | 301 | 310 | 274 | 245 | **243** | 290.100 |
| | 7 | 458 | 477 | 452 | 678 | 466 | 438 | 441 | 434 | **432** | 436 | 442 | 438 | 437 | 445.150 |
| | 8 | 548 | 583 | 555 | 756 | 575 | 517 | **515** | **515** | **515** | **515** | 530 | 524 | 523 | 532.350 |
| | 9 | 633 | 684 | 637 | 1043 | 667 | 591 | 598 | **590** | 594 | 597 | 609 | 607 | 605 | 610.300 |
| | 10 | 714 | 773 | 735 | 1118 | 761 | **667** | **667** | 682 | 672 | 670 | 688 | 686 | 683 | 689.400 |
| | 11 | 791 | 858 | 822 | 1372 | 851 | **735** | 747 | 778 | 741 | 753 | 762 | 766 | 751 | 774.350 |
| | 12 | 850 | 938 | 900 | 1449 | 929 | **802** | 809 | 880 | 806 | 809 | 814 | 856 | 854 | 865.500 |
| 6 | 7 | 1089 | 921 | 1015 | 1201 | 921 | **729** | 977 | 973 | 963 | 977 | 944 | 836 | **729** | 922.125 |
| | 8 | 1466 | 1515 | 1455 | 1763 | 1493 | 1409 | 1402 | 1401 | **1399** | 1402 | 1424 | 1416 | 1413 | 1427.55 |
| | 9 | 1840 | 1931 | 1818 | 2526 | 1889 | 1682 | 1684 | 1689 | **1681** | 1684 | 1756 | 1733 | 1731 | 1741.00 |
| | 10 | 2160 | >day | 2165 | 2834 | 2262 | **1972** | 1980 | 2027 | 1980 | 1991 | 2055 | 2038 | 2019 | 2028.00 |
| | 11 | 2459 | >day | 2496 | 3886 | 2607 | **2250** | 2255 | 2298 | 2258 | 2255 | 2261 | 2254 | 2251 | 2261.00 |
| | 12 | 2757 | >day | 2815 | 4087 | 3649 | **2512** | 2528 | 2638 | 2558 | 2528 | 2571 | 2543 | 2523 | 2532.00 |

Table 2. Test set size for CA (N; $t$, $v^7$) with $2 \leq t \leq 6$ and $2 \leq v \leq 5$.

| $t$ | $v$ | Pure computation strategies | | | | | AI-based strategies | | | | | | | | |
|---|---|---|---|---|---|---|---|---|---|---|---|---|---|---|---|
| | | Jenny Best | TConfig Best | PICT Best | IPOG-D Best | IPOG Best | DPSO Best | PSTG Best | CS Best | GS Best | HSS Best | ABCS Best | HABC Best | HABCS*m* Best | HABCS*m* Avg.Size |
| 2 | 2 | 8 | 7 | 7 | 8 | 8 | 7 | **6** | **6** | **6** | 7 | 7 | 7 | 7 | 7.40000 |
| | 3 | 16 | 15 | 16 | 15 | 17 | **14** | 15 | 15 | **14** | **14** | 15 | 15 | **14** | 15.0000 |
| | 4 | 28 | 28 | 27 | 32 | 28 | **24** | 26 | 25 | **24** | 25 | 25 | 25 | **24** | 26.1000 |
| | 5 | 37 | 40 | 40 | 45 | 42 | **34** | 37 | 37 | 36 | 35 | 38 | 37 | **34** | 37.6500 |
| 3 | 2 | 14 | 16 | 15 | 14 | 19 | 15 | 13 | **12** | **12** | **12** | 14 | 14 | 13 | 15.4500 |
| | 3 | 54 | 55 | 51 | 50 | 57 | 48 | 50 | 49 | 49 | 50 | 49 | 47 | **46** | 51.2000 |
| | 4 | 124 | 112 | 124 | 114 | 208 | 112 | 116 | 117 | 116 | 121 | 119 | 114 | **110** | 117.850 |
| | 5 | 236 | 239 | 241 | 252 | 275 | **216** | 225 | 223 | 221 | 223 | 231 | 226 | 221 | 225.475 |
| 4 | 2 | 31 | 36 | 32 | 40 | 48 | 34 | 29 | **27** | **27** | 29 | **27** | 29 | **27** | 31.8000 |
| | 3 | 169 | 166 | 168 | 226 | 185 | 150 | 155 | 155 | 153 | 155 | 157 | 155 | **149** | 158.250 |
| | 4 | 517 | 568 | 529 | 704 | 509 | **472** | 487 | 487 | 486 | 500 | 498 | 488 | 484 | 494.500 |
| | 5 | 1248 | 1320 | 1279 | 1858 | 1349 | **1148** | 1176 | 1171 | 1173 | 1174 | 1215 | 1187 | 1179 | 1196.55 |
| 5 | 2 | 57 | 56 | 57 | 80 | 128 | 59 | 53 | 53 | **51** | 53 | 54 | 54 | 55 | 57.6500 |
| | 3 | 458 | 477 | 452 | 678 | 608 | 438 | 441 | 439 | **432** | 437 | 439 | 441 | 439 | 445.350 |
| | 4 | 1938 | 1792 | 1933 | 2816 | 2560 | **1814** | 1826 | 1845 | 1821 | 1831 | 1878 | 1856 | 1855 | 1868.20 |
| | 5 | 5895 | N/A | 5814 | 9198 | 8091 | **5400** | 5474 | 5479 | 5463 | 5468 | 5630 | 5601 | 5588 | 5608.95 |
| 6 | 2 | 87 | **64** | 72 | 96 | **64** | **64** | **64** | 66 | 65 | **64** | 65 | **64** | **64** | 64.0000 |
| | 3 | 1087 | 921 | 1015 | 1201 | 1281 | **729** | 977 | 973 | 963 | 916 | 893 | 810 | **792** | 744.400 |
| | 4 | 6127 | N/A | 5847 | 5120 | **4096** | 5223 | 5599 | 5610 | 5608 | **4096** | 5541 | 5502 | 5470 | 5557.20 |
| | 5 | 23492 | N/A | 22502 | 24808 | 28513 | **20525** | 21595 | 21597 | 21532 | 21748 | 21645 | 21592 | 21565 | 21610.2 |

Table 3. Test set size for CA (N; 4,5$^P$) with $5 \leq p \leq 10$.

| $P$ | Pure computation strategies | | | | | AI-based strategies | | | | | | | | |
|---|---|---|---|---|---|---|---|---|---|---|---|---|---|---|
| | Jenny Best | TConfig Best | PICT Best | IPOG-D Best | IPOG Best | DPSO Best | PSTG Best | CS Best | GS Best | HSS Best | ABCS Best | HABC Best | HABCS*m* Best | HABCS*m* Avg.Size |
| 5 | 837 | 773 | 810 | 1250 | 908 | NA | 779 | 776 | 769 | 751 | 758 | 759 | **750** | 771.150 |
| 6 | 1074 | 1092 | 1072 | 1250 | 1239 | NA | 1001 | 991 | **984** | 990 | 1010 | 1000 | 996 | 1007.30 |
| 7 | 1248 | 1320 | 1279 | 1858 | 1349 | NA | 1209 | 1200 | **1176** | 1186 | 1198 | 1189 | 1179 | 1196.55 |
| 8 | 1424 | 1532 | 1468 | 1858 | 1792 | NA | 1417 | 1415 | 1371 | 1358 | 1413 | 1386 | **1354** | 1360.00 |
| 9 | 1578 | 1724 | 1643 | 2110 | 1793 | NA | 1570 | 1562 | 1548 | 1530 | 1621 | 1591 | **1526** | 1531.40 |
| 10 | 1719 | 1878 | 1812 | 2110 | 1965 | NA | 1716 | 1731 | 1638 | **1624** | 1831 | 1798 | 1718 | 1721.50 |

Table 4. Test size for VSCA (N, 2, $3^{15}$, {C}).

| {C} | Pure computation strategies | | | | | | AI-based strategies | | | | | | | | | |
|---|---|---|---|---|---|---|---|---|---|---|---|---|---|---|---|---|
| | WHITCH Best | TVG Best | ParaOrder Best | IPOG Best | Density Best | PICT Best | SA Best | PSTG Best | GS Best | ACS Best | PwiseGen-VSCA Best | HSS Best | ABCS Best | HABC Best | HABCSm Best | HABCSm Avg.Size |
| ∅ | 31 | 22 | 33 | 21 | 21 | 35 | **16** | 19 | 19 | 19 | **16** | 20 | 20 | 20 | 19 | 21.300 |
| CA (3, $3^3$) | 48 | **27** | **27** | **27** | 28 | 81 | **27** | **27** | 28 | **27** | **27** | **27** | **27** | **27** | **27** | 27.850 |
| CA (3, $3^4$) | 59 | 35 | **27** | 39 | 32 | 105 | **27** | 30 | 29 | **27** | **27** | 27 | 32 | 30 | 29 | 34.550 |
| CA (3, $3^5$) | 62 | 41 | 45 | 39 | 40 | 131 | **33** | 38 | 38 | 38 | **33** | 38 | 41 | 39 | 39 | 42.550 |
| CA (4, $3^4$) | 103 | **81** | NA | **81** | NA | 245 | NA | **81** | **81** | NA | **81** | **81** | **81** | **81** | **81** | 81.100 |
| CA (4, $3^5$) | 118 | 103 | NA | 122 | NA | 301 | NA | 97 | 92 | NA | 91 | 94 | **90** | 93 | **90** | 102.55 |
| CA (4, $3^7$) | 189 | 168 | NA | 181 | NA | 505 | NA | 158 | 155 | NA | 158 | 159 | 154 | 154 | **153** | 159.65 |
| CA (5, $3^5$) | 261 | **243** | NA | **243** | NA | 730 | NA | **243** | **243** | NA | **243** | **243** | **243** | **243** | **243** | 243.10 |
| CA (5, $3^7$) | 481 | 462 | NA | 581 | NA | 1356 | NA | 441 | 441 | NA | 441 | 441 | 446 | 439 | **436** | 446.15 |
| CA (6, $3^6$) | 745 | **729** | NA | **729** | NA | 2187 | NA | **729** | **729** | NA | **729** | **729** | **729** | **729** | **729** | 729.00 |
| CA (6, $3^7$) | 1050 | 1028 | NA | 967 | NA | 3045 | NA | 966 | 960 | NA | NA | 902 | 956 | 898 | **830** | 961.10 |
| CA (3, $3^4$) CA (3, $3^5$) CA (3, $3^6$) | 114 | 53 | 44 | 51 | 46 | 1376 | **34** | 45 | NA | 40 | NA | 45 | 82 | 82 | 82 | 86.000 |
| CA (3, $3^6$) | 61 | 48 | 49 | 53 | 46 | 146 | **34** | 45 | 46 | 45 | 40 | 45 | 45 | 45 | 45 | 47.55 |
| CA (3, $3^7$) | 68 | 54 | 54 | 58 | 53 | 154 | **41** | 49 | 50 | 48 | 47 | 51 | 50 | 50 | 50 | 52.20 |
| CA (3, $3^9$) | 94 | 62 | 62 | 65 | 60 | 177 | **50** | 57 | 57 | 57 | 57 | 62 | 58 | 60 | 59 | 60.20 |
| CA (3, $3^{15}$) | 132 | 81 | 82 | NS | 70 | 83 | **67** | 74 | 75 | 76 | 74 | 77 | 81 | 81 | 80 | 82.40 |

Table 5. Test size for VSCA (N, 2, $4^3\,5^3\,6^2$, {C}).

| {C} | Pure computation strategies | | | | | | AI-based strategies | | | | | | | | | |
|---|---|---|---|---|---|---|---|---|---|---|---|---|---|---|---|---|
| | WHITCH Best | TVG Best | ParaOrder Best | IPOG Best | Density Best | PICT Best | SA Best | PSTG Best | GS Best | ACS Best | PwiseGen-VSCA Best | HSS Best | ABCS Best | HABC Best | HABCSm Best | HABCSm Avg.Size |
| ∅ | 48 | 44 | 49 | 43 | 41 | 43 | **36** | 42 | NA | 41 | NA | **42** | 44 | 42 | 42 | 45.15 |
| CA (3, $4^3$) | 97 | 67 | **64** | 83 | **64** | 384 | **64** | **64** | NA | **64** | NA | **64** | **64** | **64** | **64** | 64.000 |
| MCA (3, $4^3\,5^2$) | 164 | 132 | 141 | 147 | 131 | 781 | **100** | 124 | NA | 104 | NA | 116 | 128 | 123 | 121 | 130.90 |
| CA (3, $5^3$) | 145 | **125** | 126 | 136 | **125** | 750 | **125** | **125** | NA | **125** | NA | **125** | **125** | **125** | **125** | 125.00 |
| MCA (4, $4^3\,5^1$) | 354 | **320** | NA | 329 | NA | 1920 | NS | **320** | NA | NS | NA | **320** | **320** | **320** | **320** | 320.00 |
| MCA (5, $4^3\,5^2$) | 1639 | **1600** | NA | 1602 | NA | 9600 | NS | **1600** | NA | NS | NA | **1600** | **1600** | **1600** | **1600** | 1600.0 |
| CA (3, $4^3$) CA (3, $5^3$) | 194 | **125** | 129 | 136 | **125** | 8000 | 125 | **125** | NA | **125** | NA | **125** | **125** | **125** | **125** | 125.00 |
| MCA (4, $4^3\,5^1$) MCA (4, $5^2\,6^2$) | 1220 | **900** | NA | **900** | NA | 288,000 | NS | **900** | NA | NS | NA | **900** | **900** | **900** | **900** | 900.00 |
| MCA (3, $4^3$) MCA (4, $5^3\,6^1$) | 819 | **750** | NA | **750** | NA | 48,000 | NS | **750** | NA | NS | NA | **750** | **750** | **750** | **750** | 750.00 |
| MCA (3, $4^3$) MCA (5, $5^3\,6^2$) | 4569 | **4500** | NA | **4500** | NA | 288,000 | NS | **4500** | NA | NS | NA | **4500** | **4500** | **4500** | **4500** | 4500.0 |
| MCA (4, $4^3\,5^2$) | 510 | 496 | NA | 512 | NA | 2874 | NS | 472 | NA | NS | NA | 453 | 463 | 464 | **461** | 476.75 |
| MCA (5, $4^3\,5^2$) | 2520 | 2592 | NA | 2763 | NA | 15,048 | NS | 2430 | NA | NS | NA | 2430 | 2403 | **1600** | **1600** | 1600.0 |
| MCA (3, $4^3\,5^3\,6^1$) | 254 | 237 | 247 | 215 | 207 | 1266 | **171** | 206 | NA | 201 | NA | 212 | 213 | 212 | 210 | 216.50 |
| MCA (3, $5^1\,6^2$) | 188 | **180** | **180** | **180** | **180** | 900 | **180** | **180** | NA | **180** | NA | **180** | **180** | **180** | **180** | 180.0 |
| MCA (3, $4^3\,5^3\,6^2$) | 312 | 302 | 307 | NS | 256 | 261 | **214** | 260 | NA | 255 | NA | 263 | 266 | 269 | 263 | 268.95 |

Table 6. Test size for VSCA (N, 2, $10^1 9^1 8^1 7^1 6^1 5^1 4^1 3^1 2^1$, {C}).

| {C} | Pure computation strategies | | | | | | AI-based strategies | | | | | | | | | |
|---|---|---|---|---|---|---|---|---|---|---|---|---|---|---|---|---|
| | WHITCH Best | PICT Best | IPOG Best | Density Best | TVG Best | ParaOrder Best | PSTG Best | ACS Best | HSS Best | SA Best | PwiseGen-VSCA Best | GS Best | ABCS Best | HABC Best | HABCSm Best | HABCSm Avg.Size |
| Ø | N/A | 102 | **91** | N/A | 99 | N/A | 97 | N/A | 94 | N/A | 92 | N/A | 95 | 93 | 93 | 95.25 |
| MCA (N,3, $10^1 9^1 8^1$) | N/A | 31256 | **720** | N/A | **720** | N/A | **720** | N/A | **720** | N/A | **720** | N/A | **720** | **720** | **720** | 720.00 |
| MCA (N,3, $7^1 6^1 5^1$) | N/A | 19515 | 221 | N/A | **210** | N/A | **210** | N/A | **210** | N/A | **210** | N/A | **210** | **210** | **210** | 210.00 |
| MCA (N,3, $4^1 3^1 2^1$) | N/A | 2397 | **91** | N/A | 99 | N/A | 97 | N/A | 94 | N/A | 92 | N/A | 94 | 94 | 93 | 96.500 |
| MCA (N,3, $10^1 9^1 8^1 7^1$) | N/A | 22878 | 772 | N/A | 784 | N/A | 742 | N/A | **740** | N/A | **740** | N/A | 751 | 748 | 746 | 756.45 |
| MCA (N,3, $10^1 9^1 8^1$), MCA (N,3,$7^1 6^1 5^1$) | N/A | N/A | **720** | N/A | **720** | N/A | **720** | N/A | **720** | N/A | 740 | N/A | **720** | **720** | **720** | 720.00 |
| MCA (N,3, $10^1 9^1 8^1$), MCA (N,6,$7^1 6^1 5^1 4^1 3^1 2^1$) | N/A | N/A | 5041 | N/A | **5040** | N/A | **5040** | N/A | **5040** | N/A | N/A | N/A | **5040** | **5040** | **5040** | 5040.0 |
| MCA (N,3, $10^1 9^1 8^1$), MCA (N,3,$7^1 6^1 5^1$), MCA (N,3, $4^1 3^1 2^1$) | N/A | N/A | **720** | N/A | **720** | N/A | **720** | N/A | **720** | N/A | **720** | N/A | **720** | **720** | **720** | 720.00 |
| MCA (N,4,$5^1 4^1 3^1 2^1$) | N/A | 1200 | 142 | N/A | 123 | N/A | **120** | N/A | **120** | N/A | **120** | N/A | 121 | 121 | **120** | 123.20 |
| MCA (N,5,$10^1 9^1 4^1 3^1 2^1$) | N/A | 124157 | **2160** | N/A | **2160** | N/A | **2160** | N/A | **2160** | N/A | **2160** | N/A | **2160** | **2160** | **2160** | 2160.0 |
| MCA (N,6,$7^1 6^1 5^1 4^1 3^1 2^1$) | N/A | N/A | 5041 | N/A | **5040** | N/A | **5040** | N/A | **5040** | N/A | **5040** | N/A | **5040** | **5040** | **5040** | 5040.0 |

Table 7. Test size for VSCA (N, 2, $3^{30} 10^2$, {C}).

| {C} | Pure computation strategies | | | | | | AI-based strategies | | | | | | | | | |
|---|---|---|---|---|---|---|---|---|---|---|---|---|---|---|---|---|
| | WHITCH Best | PICT Best | IPOG Best | Density Best | TVG Best | ParaOrder Best | SA Best | HSS Best | PSTG Best | GS Best | ACS Best | PwiseGen-VSCA Best | ABCS Best | HABC Best | HABCSm Best | HABCSm Avg.Size |
| Ø | N/A | **100** | 101 | N/A | 101 | N/A | **100** | 106 | 102 | N/A | **100** | N/A | 107 | 105 | 105 | 109.10 |
| CA (N,3, $3^{20}$) | N/A | 940 | **100** | N/A | 103 | N/A | **100** | 109 | 105 | N/A | **100** | N/A | 120 | 118 | 117 | 122.10 |
| MCA (N,3, $3^{20} 10^2$) | N/A | 423 | N/S | N/A | 423 | N/A | **304** | 450 | 481 | N/A | **304** | N/A | 504 | 503 | 503 | 507.85 |
| MCA (N,4, $3^3 10^1$) | N/A | 810 | **273** | N/A | **270** | N/A | N/A | **270** | **270** | N/A | N/A | N/A | **270** | **270** | **270** | 270.75 |
| MCA (N,5, $3^3 10^2$) | N/A | **2800** | **2700** | N/A | **2700** | N/A | N/A | **2700** | **2700** | N/A | N/A | N/A | **2700** | **2700** | **2700** | 2700.0 |
| MCA (N,6, $3^4 10^2$) | N/A | N/A | **8100** | N/A | **8100** | N/A | N/A | **8100** | **8100** | N/A | N/A | N/A | **8100** | **8100** | **8100** | 8100.0 |

Table 2 depicts experimental set 2, where the HABCS*m* manages to get the best test set sizes (or most minimum 9 out of 20 entries). Note that 3 out of the 9 best result entries are new minimal test set sizes obtained by the HABCS*m*. However, DPSO surpasses all of the other existing strategies in achieving the best test set sizes (or most minimum 11 out of 20 entries). The other strategies shared the best results for less than 5 entries. Similar to the observations in Table 1, TConfig and PICT as well as Jenny and IPOG-D, have not reported significant or minimum test set size for any entries.

In experimental set 3, shown in Table 3, the HABCS*m*, GS and HSS achieve the best test set sizes overall. Clearly, the HABCS*m* achieves 50 % of the best sizes (3 out of 6 entries). Additionally, the best sizes that were achieved are the new best sizes generated by the HABCS*m* for the system configurations (P = 5, 8 and 9 with 750, 1354 and 1526 test cases, respectively). The GS retains the best sizes for 2 test configurations (P = 6 and 7, recording 984 and 1176 test cases, respectively). The HSS only gives the best results for one entry (P = 10 with 1624 test cases). For experimental set 3, only 3 strategies were able to achieve the best sizes. The other strategies produce acceptable results as compared to the best sizes. Whereas DPSO have been not reported any results in the literature.

Referring to the Variable Strength Covering Array (VSCA) benchmarked experimental sets 4, 5, 6 and 7, It is generally observed that probabilistic based strategies outperform deterministic based strategies for both main-strength and sub-strength generations. Based on experimental set 4 in Table 4, the HABCS*m* achieves 50 % of the best sizes obtained (8 out of 16 entries). The HABCS*m* manages to obtain three new minimal test set sizes (in the following sub-configurations: CA (S, 4, $3^7$), CA (S, 5, $3^7$), and CA (S, 6, $3^7$)). In these test configurations, SA generates the best test set in general (9 out of 16 entries). Specifically, the SA in the low interaction strength achieved the best results as it only supports *t* up to 3 ($t \leq 3$). The PWiseGen-VSCA, ABCS, HSS, HABC and PSTG produce competitive results as well with 7, 5, 4, 4 and 4 best sizes. The PICT, WHITCH and Density generates the worst results as it only supports *t* up to 3 ($t \leq 3$) for the variable-strength test configuration.

Table 5 demonstrates that in experimental set 5, the HABCS*m* outperformed other existing strategies to achieve the best results. The best result achieved was for 11 out of 15 entries. The HABCS*m* manages to obtain a new minimal test set size in the case of MCA (S, 4, $4^3 5^2$) sub-strength configuration with 461 test cases. HABC and HSS perform well to get the best result for 10 out of 15 entries each. Whereas ABCS and PSTG generate the best results for 9 out of 15 entries each. The other strategies also generated competitive results when compared with the best results. The WHITCH and PICT consistently produce the worst results for the test configuration. However, GS and PwiseGen-VSCA have not reported any results in the literature.

Experimental set 6, presented in Table 6, demonstrates acceptable performance by several strategies (e.g., HABCS*m*, ABCS, HABC, HSS, PSTG, PWiseGen-VSCA, TVG and IPOG) for this test generation. The HABCS*m* and PSTG excel in most of the best results (8 out of 11 entries each). Regarding HSS, ABCS, HABC and PWiseGen-VSCA, these strategies perform equally well, demonstrating the best test set sizes (7 out of 11 entries each). In the same manner, the IPOG generates competitive test set size in many sub-strength configurations of the best sizes. The PICT generated the poorest results in most cases with no best size obtained among all the test configurations (with some missing results). As for SA, Density, ParaOrder, WHITCH, GS and ACS, no published results are available.

Concerning the experimental set 7 in Table 7, the HABCS*m*, ABCS, HABC, HSS, PSTG, SA, IPOG, ACS and TVG are each able to obtain the best sizes (3 out of 6 entries). In fact, the HABCS*m*, ABCS, HABC, HSS, TVG, IPOG and PSTG generate the optimal test set size for high interaction strength (t > 3). However, the SA and ACS dominate in low interaction strength (t ≤ 3). The PICT strategy obtains the best sizes (1 out of 6 entries). Regarding Density, ParaOrder, WHITCH, GS and PwiseGen-VSCA, no published results are available.

## 7. Conclusion and Future work

This paper proposes a new Meta-heuristic based *t*-way strategy called Hybrid Artificial Bee Colony strategy (HABCS*m*) based on merging the advantages of the Artificial Bee Colony (ABC) algorithm with a Particle Swarm Optimization (PSO) algorithm to improve the performance of the original ABC. HABCS*m* is the first *t*-way strategy that adopts HABC algorithm as its core implementation for generating a final test set and also adopts the Hamming distance as the final selection criterion enhancing the exploration of new solutions. The experiments that were conducted have shown that HABCS*m* has a better performance for generating the best test set size than existing strategies, as well as variants of ABC. In many entries, HABCS*m* outperformed the other strategies. However, if HABCS*m* failed to generate the optimum results, the generated results were within reasonable values. The results of the experiment were encouraging. Thus, we are planning to adopt HABCS*m* to complete implementation to support automated test execution and other *t*-way test generation types. In particular, several *t*-way features needed to be included (i.e. input-output relations *t*-way, sequencing *t*-way and constraint's *t*-way).

*Cite this paper as:*

Alazzawi K. Rais H.M.; Basri S.; Alsariera YA; Capretz L.F.; Imam A.A.; Balogun A.O.; HABCSm: A Hamming Based *t*-way Strategy Based on Hybrid Artificial Bee Colony for Variable Strength Test Sets Generation, *International Journal of Computers Communications* & *Control*, 16(5):1-18, #4308, October 2021.
https://doi.org/10.15837/ijccc.2021.5.4308